\documentclass
[aps,
twocolumn,
superscriptaddress,
]{revtex4-2}

\usepackage{graphicx}
\usepackage{xcolor}
\usepackage[colorlinks=true,citecolor=blue,linkcolor=magenta]{hyperref}

\usepackage[utf8]{inputenc}
\usepackage{amsmath,bm}
\usepackage{subfigure}

\begin{document}

\title{Unlocking the power of global quantum gates with machine learning}

\author{Vinit Singh}
\affiliation{Los Alamos National Laboratory, Theoretical Division, Los Alamos, New Mexico 87545, USA}
\affiliation{Purdue University, Department of Chemistry, West Lafayette, Indiana 47906, USA}
\affiliation{North Carolina State University, Department of Electrical and Computer Engineering, Raleigh, NC 27606, USA}

\author{Bin Yan}
\affiliation{Los Alamos National Laboratory, Theoretical Division, Los Alamos, New Mexico 87545, USA}

\begin{abstract}
In conventional circuit-based quantum computing architectures, the standard gate set includes arbitrary single-qubit rotations and two-qubit entangling gates. This choice is not always aligned with the native operations available in certain hardware, where the natural entangling gates are not restricted to two qubits but can act on multiple, or even all, qubits simultaneously. However, leveraging the capabilities of global quantum operations for algorithm implementations is highly challenging, as directly compiling local gate sequences into global gates usually gives rise to a quantum circuit that is more complex than the original one. Here, we circumvent this difficulty using a variational approach. Specifically, we study parameterized circuit ansatze composed of a finite number of global gates and layers of single-qubit unitaries. We demonstrate the expressibility of these ansatze and apply them to the problem of ground state preparation for the Heisenberg model and the toric code Hamiltonian, highlighting their potential for offering practical advantages.
\end{abstract}

\flushbottom
\maketitle
\thispagestyle{empty}

Implementing arbitrary quantum circuits on hardware requires selecting a universal gate set compatible with the underlying physical architecture. The most common choice comprises arbitrary single-qubit gates and two-qubit CNOT gates. Known decompositions, such as the Solovay-Kitaev and Cartan decompositions~\cite{dawson2005solovay, kitaev2002classical, PhysRevA.108.052607}, allow the transpilation of any unitary operation into this gate set. While arbitrary single-qubit gates are native to most quantum hardware, CNOT gates often require decomposition into alternative physically implementable entangling gates like the Controlled-Z, Echoed Cross Resonance, or Mølmer-Sørensen gates, depending on the device's physical platform.

This mismatch between the low-level quantum hardware's native capabilities and high-level quantum algorithms becomes particularly evident in systems like trapped ions, where the natural entangling operations can act on more than two qubits simultaneously, or even globally across all qubits~\cite{nigg2014quantum, grzesiak2020efficient, bassler2023synthesis}. Using intermediate CNOT gates in such architectures is inefficient, requiring additional overhead for their decomposition~\cite{van2021constructing}. Moreover, this approach forfeits the advantages of native multi-qubit operations, which can apply layers of commuting two-qubit interactions simultaneously across the system.

Entangling gates, being noisier and more resource-intensive than single-qubit gates, often dominate the computational cost on quantum devices. Thus, minimizing their use can significantly improve both efficiency and fidelity. Recent studies suggest that global gates, quantum gates based on global control, can efficiently represent a broad class of circuits \cite{maslov2018use}, and can in fact represent any quantum circuit due to their universality. High-fidelity global entangling operations were also experimentally demonstrated~\cite{Landsman2019,Figgatt2019,Nigg2014,grzesiak2020efficient,debnath2016demonstration, martinez2016compiling, cesa2023universal}. Despite considerable efforts, the literature on synthesizing circuits with global gates for practical algorithm implementations remains relatively sparse~\cite{Sinitsyn2023Topologically}. Whether global gates can offer potential advantages for practical tasks remains exclusive.

\begin{figure}[b!]
    \centering
    \includegraphics[width=0.9\linewidth]{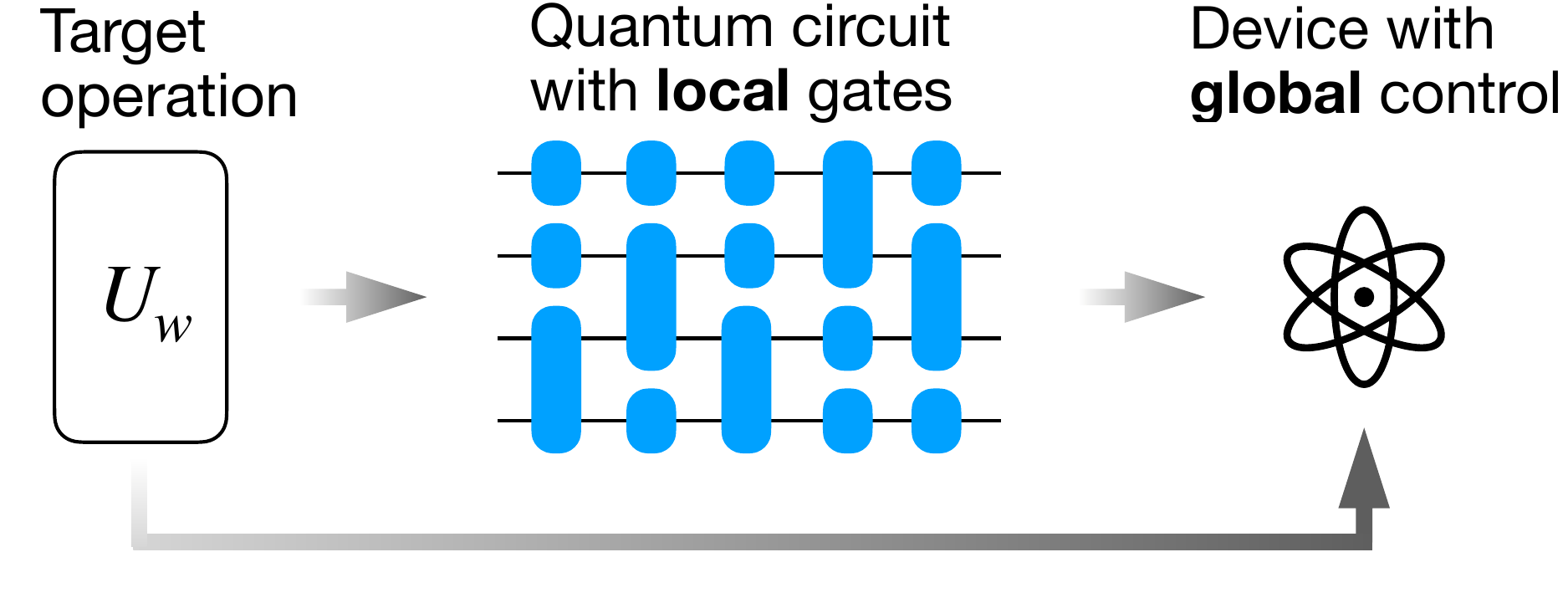}
    \caption{To implement a desired operation $U_\omega$ (e.g., a component of a quantum algorithm), quantum computing with \emph{local} gates undergoes two layers of compilations. That is, compiling to local-gate sequences and compiling to hardware implementations. Our proposal aims at achieving the same operation directly with the underlying global control capabilities (longer arrows) using a variational approach.}
    \label{fig:illustration}
\end{figure}

In this work, we address this gap with a variational approach (Figure~\ref{fig:illustration}). Specifically, we implement variational quantum algorithms using circuit ansatzes that are directly constructed from alternating layers of global gates and single-qubit rotations. This design restricts the circuit to a constant number of global gates, ensuring that the implementations remain low-cost and are suitable for near-term quantum computers supporting such gates. Despite these constraints, we demonstrate that the ansatzes are highly expressive and capable of preparing a diverse range of quantum states, including those with long-range entanglement.

\section{Global Gates}

Global gates are unitary operations that simultaneously act on multiple or all qubits in a system. They are native to several platforms and have been demonstrated experimentally across a range of hardware~\cite{Landsman2019,Figgatt2019,Nigg2014,grzesiak2020efficient,debnath2016demonstration, martinez2016compiling, cesa2023universal, menta2024globally, cioni2024conveyor, shapira2023fast, shapira2020theory, shapira2025programmable, schwerdt2024scalable}. They are typically achieved by driving the systems through their native underlying interactions and with global control pulses. For example, in Rydberg atom systems, global operations can be implemented using resonant laser pulses over a static atomic arrangement~\cite{cesa2023universal}. Superconducting systems can emulate global gates by coupling the Josephson junction qubits to a shared resonator mode~\cite{devoret2013superconducting}. 

For concordance, in this work, we choose to work with a global entangling gate native to trapped ion systems, where the ion qubits couple through Mølmer-Sørensen interaction~\cite{sorensen1999quantum}. This, together with globally dressing amplitude-modulate laser pulses, achieves Global Control $Z$ (GCZ) gates with specific rotation angles for any target subsets $\mathcal{S}$ of qubits with arbitrary pairwise connectivity~\cite{van2021constructing, grzesiak2020efficient, figgatt2019parallel}, i.e.,
\begin{equation}\label{eq:GCZ}
    {\rm GCZ}(\vec{\theta}) = \prod_{i \sim j\in \mathcal{S}} {\rm CZ}_{ij} (\theta_{ij}).
\end{equation}
Here, $i\sim j$ indicates qubit $i$ and $j$ are connected;
\begin{equation}\label{eq:CZ}
    {\rm CZ}(\theta) = e^{i \theta |11\rangle \langle 11| }
\end{equation}
is the parameterized two-qubit control $Z$ gate (CZ), and we assume the angle $\theta$ can be tuned individually for each pair of qubits. It is worth stressing that finding the desired pulses for realizing the above gate generally takes polynomial time in the total number of qubits~\cite{grzesiak2020efficient}, which is not much more than that for a single two-qubit control $Z$ gate~(\ref{eq:CZ}). 

In addition, as single-qubit gates can be implemented efficiently in high fidelity with a small execution time compared to the entangling operations on the physical device, we treat single-qubit gates as a free resource. This also makes working with a variety of entangling gates equivalent so long as they can be obtained from the GCZ gates using a simple single-qubit basis change. For instance, we treat the GCX gate
\begin{equation}\label{eq:GCX}
    {\rm GCX}(\vec{\theta}) = \prod_{i \sim j\in \mathcal{S}} {\rm CX}_{ij} (\theta_{ij})
\end{equation}
as an element in our global gate set. Here the two-qubit control $X$ gate ${\rm CX}(\theta)$ is defined similarly.

\subsection{the Power of Global Gates}

Global gates, combined with arbitrary single-qubit operations, can form a universal gate set, making them capable of realizing any quantum circuits. Moreover, they allow efficient synthesis of certain classes of quantum circuits that would require significantly more resources using conventional two-qubit entangling gates. For instance, any \(n\)-qubit Clifford operation can be implemented with a constant number of global gates, independent of the system size, as opposed to $O(n^2/\ln(n))$ gates in a local-gate approach~\cite{bravyi2022constant}. An arbitrary circuit can be decomposed into Clifford components and single-qubit rotations. This simplifies the synthesis of common circuits like QFT~\cite{maslov2018use} and exponentiated Pauli gates~\cite{van2021constructing}, crucial in applications such as Trotterized quantum chemistry. Furthermore, their experimental feasibility, with pulse complexity comparable to two-qubit gates, establishes them as an efficient and practical tool for quantum circuit design.

Despite these appealing features, approaches for leveraging the advantages of global gates to fully unleash their potential in quantum algorithm implementation remain poorly investigated. The difficulties stem from two major challenges: i) Directly compiling local gate quantum circuits into global gates usually does not offer any advantages. For instance, for circuits composed of CNOT gates separated by non-Clifford single qubit gates, naively replacing each CNOT with $26$~\cite{bravyi2022constant} (albeit finite) GCZ gates would make the physical implementation of the circuit even more complex; ii) A global quantum gate---a unitary matrix acting on the entire Hilbert space of all the qubits---consists of a vast number of parameters that makes it nearly impossible to directly design quantum algorithm out of them.

In this work, we propose a novel approach that extends the applicability of global gates. We construct parameterized quantum neural networks (variational ansatze) composed of only a finite number of global gates (GCZ or GCX) with single qubit unitaries. These ansatze are specifically structured to avoid the issue of barren plateaus (discussed in the following section), thus making them efficiently trainable and meanwhile maintaining their high expressibility.

\section{Varitional Approach}

In this section, we describe the method in detail. We focus on the problem of ground state preparation for given Hamiltonians, though the proposed approach is broadly applicable to various other problems, such as unitary compiling. 

The problem is formulated as an optimization task, where a cost function is minimized to quantify the accuracy of the simulated state. To achieve this, we employ a variational approach, leveraging parameterized quantum circuits (PQC) as an ansatze for the target quantum state. These circuits are designed with tunable parameters $\vec\theta$ that can be optimized during the training process. The cost function can be defined as the expectation value of the Hamiltonian, i.e.,
\begin{equation}
    C_{\rho, H}(\vec\theta) = \text{Tr}[H U(\vec\theta) \rho U(\vec\theta)^\dagger],
\end{equation}
where $\rho$ denotes the density matrix of the initial state of the system; $H$ is the Hamiltonian whose ground state is to be discovered. The PQC applies a unitary transformation $U(\vec\theta)$ to the initial state $\rho$. The expectation value of $H$ with respect to the evolved state quantifies how closely the prepared state aligns with the desired target state. The variational algorithm optimizes the parameters $\vec\theta$ to minimize $C_{\rho, H}(\vec\theta)$, thereby preparing a quantum state that minimizes the energy. The algorithm adopts a hybrid quantum-classical framework: the PQC execution is carried out on quantum hardware to compute an estimate of the cost function, while the parameter optimization is performed on a classical device.

\begin{figure*}[t!]
    \centering
    \includegraphics[width=1\linewidth]{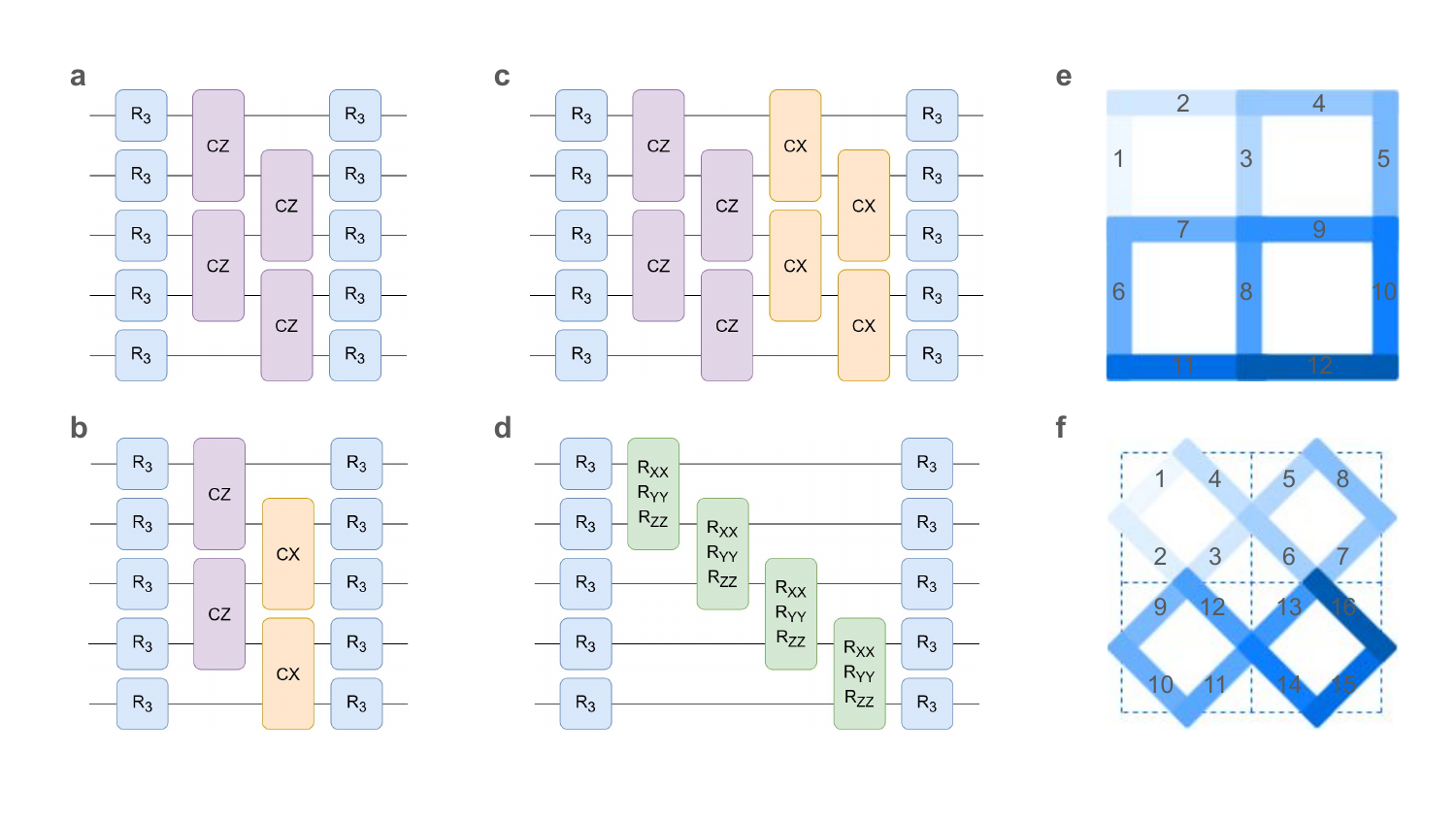}
    \caption{The figure illustrates the minimal version of each ansatz with a single layer ($k=1$). The first two columns depict the global gate ansatze: (a) GZ, (b) ${\rm GZX}_{\rm H}$, (c) GZX, and (d) Cartan, arranged in a 1D brick-wall format (equivalent to a ladder format in this case, as the $2$-qubit CZ gates commute), where the $2$-qubit gates are applied to neighboring pairs. The single-qubit rotations are defined as $R_3 = R_Z(\theta_3) R_Y(\theta_2) R_Z(\theta_1)$, where $R_Z$ and $R_Y$ are single qubit rotations along the $Z$ and $Y$ axis, respectively. For the global gate ansatze, the $2$-qubit entangling gates are either CZ or CX. For the Cartan ansatz, they are arbitrary 2-qubit unitaries decomposed into $R_{XX}$, $R_{YY}$, and $R_{ZZ}$, which are generated by operators $XX$, $YY$, and $ZZ$, respectively. 
    The rightmost column illustrates the geometry of the $2$-qubit entangling gate layer in the 2D ansatze. Each rectangular block between nearest neighbors represents an entangling gate applied to the corresponding qubit pair. The order of gate application is indicated by a color gradient, with lighter blue representing earlier gates and darker blue representing later ones. (e) shows the arrangement of entangling gates on a square lattice, where gates are applied between each nearest-neighbor pair sequentially, first sweeping from left to right and then from top to bottom. (f) demonstrates the same for the Toric code lattice. }  
    \label{fig:ansatze}
\end{figure*}

The design of ansatze is critical to the success of the protocol. To ensure effective state preparation and optimization, the PQC must satisfy two essential criteria:  
\begin{enumerate}
    \item Trainability---The circuit should avoid barren plateaus, which can impede convergence and lead to optimization stagnation into sub-optimal solutions.
    \item Expressibility---The circuit must be capable of representing the desired quantum states with high fidelity.
\end{enumerate}

We are going to discuss each of these criteria in detail and bolster how our proposed circuits fulfill them. 

\subsection{Barren Plateau}

One key requirement for any successful ansatz is its ability to train well, starting from any arbitrary initial state and for a wide variety of Hamiltonians. The gradient-based optimizers rely on sufficiently large gradient values so that the parameters of the ansatz can be updated. However, it is well known that sufficiently complex ansatze (as random as a unitary $2$-design) admits the infamous ``barren plateau'' phenomenon~\cite{Mcclean2018,Larocca2024}. That is the variance of the gradient of the cost function, ${\rm Var} \left[\partial_{\vec{\theta}} C\right]$ decays exponentially with the system size, making the landscape of the cost function essentially flat and untrainable.

Numerous methods have been adopted to surpass the barren plateau issue, with various benefits and drawbacks. Here, we focus on a restricted class of circuits with depths no more than the logarithm of the number of qubits. Such circuits are known to be free of barren plateau~\cite{Cerezo2021Cost,zhang2024absence} when the input states and final measurements are further restricted to product state and local observables, respectively. However, such circuits can be efficiently simulated classically and hence do not directly offer quantum advantages. Nevertheless, the post-trained ansatze can used as an intermediate building block for subsequential applications. The crucial question asked is whether such circuit ansatze are sufficiently expressive. This question does not have a general answer and requires case-by-case analysis~\cite{Larocca2024}. In the following, we will demonstrate the global-gate ansatze is sufficiently expressive to represent even long-range entangled states.

\subsection{Global Gate Ansatze}

Our general strategy for constructing the global gate ansatze starts from building the circuit with a \emph{finite} number ($k$) of alternating layers of single-qubit gates and two-qubit gates. See Figure~\ref{fig:ansatze} for an illustration. The two-qubit layers are designed to have finite local depth and are, therefore, guaranteed to be free of barren plateau. Crucially, each layer of two-qubit gates consists of only the same type of gates, i.e., ${\rm CZ}(\theta)$ as defined in~(\ref{eq:CZ}), such that they can be implemented using a \emph{single} global gate GCZ in equation~(\ref{eq:GCZ}).

We compare a variety of ansatze as constructed with the above general procedure. The detailed structures of the ansatze also depend on the underlying architecture (qubit connectivity). For simplicity, we first describe ansatze with qubits arranged in a $1$-D geometry. In this case, each layer of 2-qubit unitaries connects only neighboring sites in a ladder format (Figure~\ref{fig:ansatze}). We also restrict our discussion to finite-depth ansatze, though they can be easily generalized to $\mathcal{O}(\log(n))$ depth. Our numerical simulations are limited to relatively small system sizes and, therefore, cannot reveal clear distinctions between finite and log-depth in both the circuit structures and the final performance in state preparations.

\begin{itemize}
    \item $\rm{\bm{GZ}}$ \textbf{ansatz}: The ansatz consists of a layer of single-qubit unitaries and a layer of 2-qubit CZ gates connecting neighboring sites in a ladder format. Then, the above layers are repeated $k$ times. Each layer of the 2-qubit CZ gates can be viewed as a single GCZ gate. The circuit depth of this ansatz is $2k$, i.e., $k$ layers of single qubit gates and $k$ global gates.

    \item $\rm{\bm{GZX}}$ \textbf{ansatz}: The ansatz is composed of a layer of single qubits unitaries, then a layer of 2-qubit CZ gates connecting neighboring sites in a ladder format, followed by a layer of 2-qubit CX gates connecting neighboring sites in a ladder format as well. Then, the above layers are repeated $k$ times. The circuit depth of this ansatz is $3k$, i.e., $k$ layers of single qubit gates and $2k$ global gates.

    \item $\rm{\bm{GZX_{\rm H}}}$  \textbf{ansatz}: The $\rm{GZX_H}$ circuit is a variant of the previous GZX ansatz, which keeps the total number of 2-qubits gates same as the GZ ansatz, by splitting the two-qubit gate set into two groups, $ A \in \{ (k, k+1) | k~{\rm even} \} $  and $ B \in \{ (k, k+1) | k~{\rm odd} \} $. First a layer of CZ gates are acted on pairs of qubits in the A configuration, and then a layer of CX gates are acted on pairs of qubits in the B configuration.

    \item \textbf{Cartan Ansatz}: We also compare the above ansatzes, which can be implemented with a finite number of global gates, to the most general circuits with local finite depth~\cite{zhang2024absence}, where each two-qubit gate can be an arbitrary two-qubit unitary. The name is inspired from the Cartan decomposition of arbitrary 2-qubit unitaries. This ansatz cannot be efficiently realized using global gates but nevertheless serves as a benchmark for our global gate circuits.
\end{itemize}

\begin{figure*}[t!]
    \centering
    \includegraphics[width=1\linewidth]{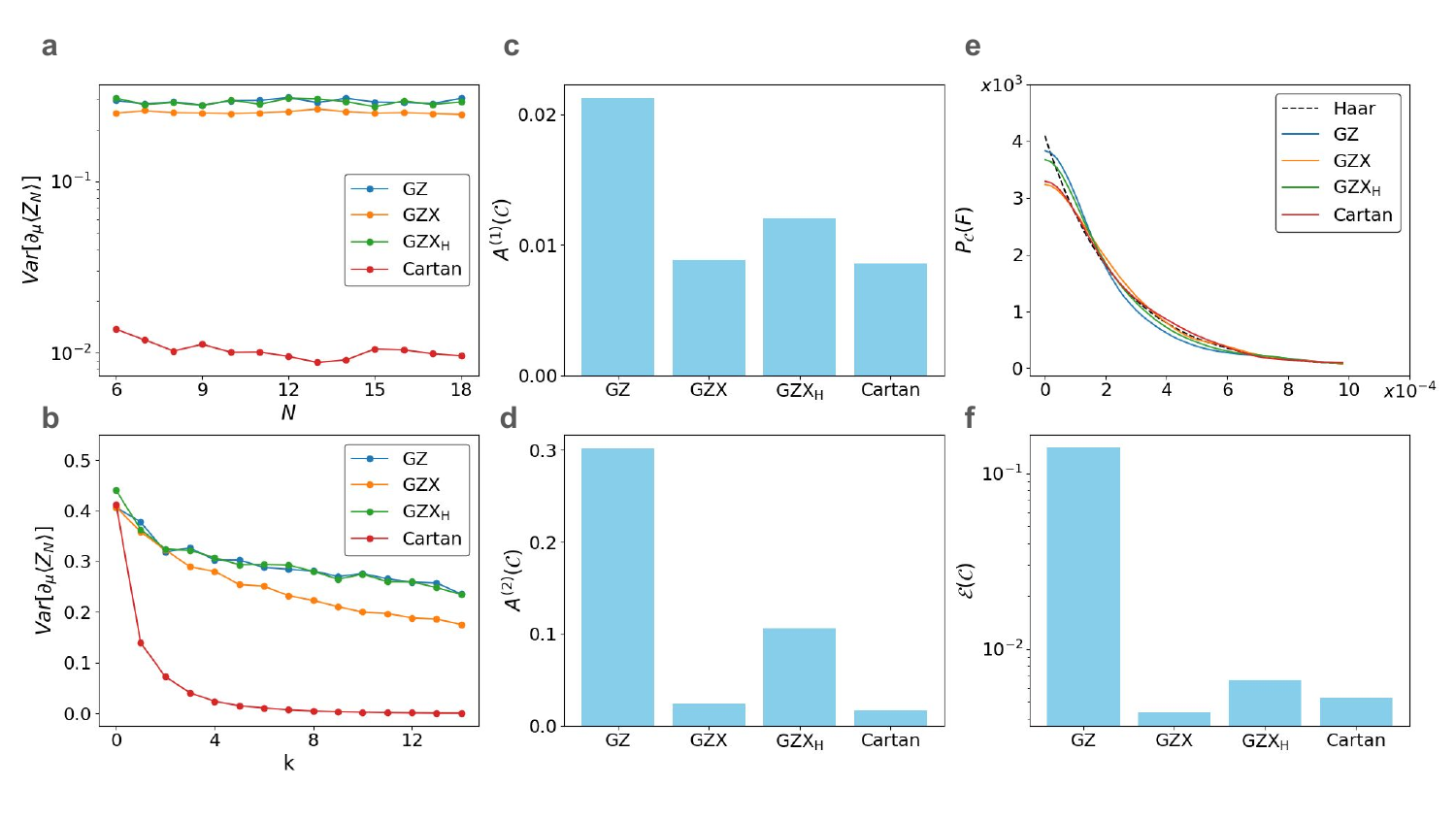}
    \caption{To quantify the trainability of the ansatze, we compute the variance of the gradient w.r.t a parameter $\mu$ for $Z_N$ Hamiltonian - a single Pauli-$Z$ operator acting on the last qubit. Here, we fix $\mu$ to be the parameter of the RY gate on the last qubit of the first $R_3$ layer in each ansatz. In the Appendix we discuss more on the variance of gradient w.r.t other parameters. (a) Upon increasing system size, the variance of the gradient, ${\rm Var}[\partial_\mu \langle Z_N \rangle]$, remains constant, indicating the absence of a barren plateau for all ansatze at $k=6$. (b) The variance ${\rm Var}[\partial_\mu \langle Z_N \rangle]$ decays exponentially as the circuit depth $k$ increases with a fixed system size $N=16$. Each variance data point is computed from $1000$ samples. In the remaining plots, we compare the expressibilities of various ansatze on a $3 \times 3$ lattice with 12 qubits that we use to train the toric code model. An ensemble of $10,000$ random circuit instances is used for the analysis. (c) and (d) shows $A^{(t)}(\mathcal{C})$ [(\ref{eq:distance})] for $t=1$ and $t=2$, respectively. Lower values of the moments indicate higher expressibility, with a perfect Haar measure yielding zero for all moments. (e) shows the probability distributions for the fidelities between two states from the ensembles generated from the ansatze, whose K-L divergences [(\ref{eq:KL})] compared against Haar measure are depicted in (f).}
    \label{fig:expressibility}
\end{figure*}

The ansatze constructed using the above procedure has finite depth and is therefore barren plateau free. We illustrate this fact with numerical simulations in Figure~\ref{fig:expressibility} (a,b).

One can generalize the 1D ansatze to higher dimensions by following a systematic order in which two-qubit gates are applied. For instance, on a square lattice, starting from the top-left lattice point $(0,0)$, a two-qubit gate is applied first to its bottom neighbor $(1,0)$ and then to its right neighbor $(0,1)$. The process continues by moving to the next lattice point to the right, applying gates to its corresponding bottom and right neighbors. Once a row is completed, the process repeats for the next row. More generally, at any lattice point $(x,y)$, gates are applied between pairs $(x,y) \leftrightarrow (x+1,y)$ and $(x,y) \leftrightarrow (x,y+1)$. If a lattice point is at an edge, any unavailable neighbors are simply skipped, and no period boundary conditions are enforced. Figure~\ref{fig:ansatze} (e) illustrates this approach, where lighter colors indicate gates applied first, followed by darker ones. The numbers on the links denote the ordering of gate applications.

To construct a 2D version of any global gate ansatze introduced in this work, we begin by applying single-qubit $R_3$ gates at every lattice point, followed by the two-qubit gates using the ordering described above. For the GZ ansatz, a single layer consists of applying all single-qubit $R_3$ gates, followed by a layer of CZ gates. The GZX ansatz follows a similar structure but with an additional step: after applying a layer of $R_3$ gates and CZ gates, an additional layer of CX gates is applied following the same procedure. The ${\rm GZX}_{\rm H}$ ansatz follows a slightly modified approach, where the links between qubits are assigned natural numbers based on their order of application and then divided into two groups: odd-numbered and even-numbered links. A single ${\rm GZX}_{\rm H}$ layer is constructed by first applying a layer of $R_3$ gates to all qubits, followed by the application of CZ gates on all odd-numbered links according to the assigned ordering. Once all CZ gates are applied, CX gates are applied on all even-numbered links in the same prescribed order. 

Again, it is important to emphasize that all CZ (or CX) gates in a layer can be implemented simultaneously as a single global gate operation. So, the GCZ ansatz requires only one global gate per layer, whereas GZX and ${\rm GZX}_{\rm H}$ ansatze require two global gates per layer.  

One of the primary lattices considered in this paper is the Toric code lattice, where qubits are positioned at the midpoints of the edges of a square lattice. We adopt a gate application order similar to that of the square lattice, moving from left to right and then top to bottom [see Figure~\ref{fig:ansatze} (f) for the lattice structure]. At each plaquette, 2-qubit gates are applied between nearest-neighbor qubits, forming a rhombus-shaped ($\diamond$) pattern. Alternatively, one could connect the bottom qubit within each plaquette to the other three neighboring qubits in the plaquette, forming a claw-shaped pattern, as shown in Figure~S80 of~\cite{zhang2024absence}. However, in this paper, we exclusively use the $\diamond$-shaped pattern. It should be noted that for both the square and toric code lattices, the local depth of the ansatze remains at most four per layer, ensuring the finite local depth.

This ansatz construction is not limited to square lattices; it can be extended to any arbitrary lattice, provided a well-defined ordering of 2-qubit gate applications is established. The pairs of sites should ideally be chosen to be local, ensuring that the circuit maintains a low local depth. For instance, in a Kagome lattice, where qubits reside at the vertices of corner-sharing triangles, gates are applied between the three qubits within each triangle. Once a set of three gates has been applied to one triangle, the process moves on to an adjacent triangle. 

\subsection{Expressibility}

In this section, we describe the metrics used to quantify the expressibility of a given quantum circuit ansatz. 

The expressibility of a circuit ansatz $\mathcal{C}$ can be quantified by comparing the state distribution generated by $\mathcal{C}$ to the Haar distribution. The idea is that a more expressive ansatz generates a state ensemble that covers the entire Hilbert space more uniformly. Therefore, the expressibility of an ansatz can be estimated by the distance between moments of its generated state ensemble to the same moment of the Haar random ensemble, i.e.,
\begin{equation}\label{eq:distance}
A^{(t)}(\mathcal{C}) =  \left\| \int dU~ (U \rho U^{\dagger})^{\otimes t} - \int d\vec{\theta}~\left[U(\vec{\theta}) \rho U(\vec{\theta})^{\dagger}\right]^{\otimes t} \right\|.
\end{equation}
Here the first integral is taken with respect to the Haar measure. If $A^{(t)}(\mathcal{C}) = 0$, the ansatz $U(\vec{\theta})$ forms a $t$-design. Smaller values of $A^{(t)}$ indicate greater expressibility. We only consider $t\le 2$ since the global gate anzates as constructed are barren plateau free, and hence cannot be more random than a unitary $2$-design.

We choose the distance metric $\|\cdot\|$ as the trace distance ($1$-norm), which is a strong distance measure with an operational interpretation~\cite{watrous2018theory}. Other distance metric can be adopted, such as the Hilbert-Schmidt distance. This distance metric results in a measure called the Frame Potential~\cite{Nakaji2021Expressibility} and will be discussed in the appendix.

We also consider a metric that directly quantifies the probability distribution of state ensemble generated by the circuit ansatze, namely, the Kullback–Leibler (KL) divergence~\cite{Nakaji2021Expressibility}. This quantity captures the difference between the probability distributions of the fidelity $F = |\langle\psi|\psi'\rangle|^2$ between two random states generated by the ansatz $\mathcal{\mathcal{C}}$ and that of the Haar distribution. It is formally defined as:
\begin{equation}\label{eq:KL}
\begin{aligned}
\mathcal{E}(\mathcal{C}) &= D_{\text{KL}} (P_\mathcal{C}( F )\|P_{\text{Haar}}(F)) \notag \\
&\equiv \int_{0}^{1} df~ P_\mathcal{C}(F) \ln \frac{P_\mathcal{C}(F)}{P_{\text{Haar}}(F)},
\end{aligned}
\end{equation}
where $P_\mathcal{C}(F)$ is the probability distribution of the fidelity $F$ for the ansatz $\mathcal{C}$, and $P_{\text{Haar}}(F)$ is the corresponding distribution for the Haar measure, which is given by the Porter-Thomas distribution,
\begin{equation}
P_{\text{Haar}}(F) = (d- 1)(1 - F)^{d-2},
\end{equation}
where $d$ is the dimension of the Hilbert space. Since for two distributions $p$ and $q$, $D_{\text{KL}}(q\|p) = 0$ if and only if $q = p$, a smaller value of $\mathcal{E}(\mathcal{C})$ indicates higher expressibility.

Figure~\ref{fig:expressibility} (c-f) depict $A^{(t)}$ and the KL divergence for various 2D ansatze for the toric code model, indicating their high expressiblities. Notably, the GZX ansatze is on par with the most general Cartan ansatze, despite that it's two-qubit gate set is more restricted and much simpler.

\section{Applications}

We have demonstrated that the proposed variational ansatze using global gates is efficiently trainable and sufficiently expressible. To test the applicability of the developed approach in real case scenarios, we apply it to the problem of ground state preparation for two important examples in many-body physics: the Heisenberg model and the Toric code Hamiltonian.

\subsection{Training Process}

We first describe the general training procedure to simulate the ground state of a given Hamiltonian $H$. We begin by choosing a variational ansatz, which will approximately represent the ground state of the target Hamiltonian post optimization. In this optimization problem, the energy serves as the cost function, which we iteratively minimize by tweaking the ansatz parameters $\vec{\theta}$ in the PQC. The energy is evaluated with respect to the output state of the PQC, $E = \langle \psi(\vec{\theta}) | H | \psi(\vec{\theta}) \rangle$. We consider local Hamiltonians with a polynomial number of local terms to ensure efficient evaluations. The optimization process requires computing gradients of energy with respect to each parameter. We employ the parameter shift rule, which allows gradient estimation by evaluating the circuit at shifted parameter values. Once the gradients are obtained, we update the parameters using gradient-based optimization algorithms. In our experiments, we use the \emph{Adam} optimizer~\cite{kingma2014adam}, a widely used adaptive gradient-based method, to iteratively refine the parameters. The optimization steps proceed as follows:

\begin{enumerate}
    \item Initialization: The parameters $\vec{\theta}$ are initialized randomly. The choice of initialization plays a crucial role in the convergence behavior, as poor initialization can lead to slow convergence or local minima.
    \item Iterative Optimization: The optimizer updates parameters in each epoch based on the computed gradients, progressively reducing the energy. These updates continue until the energy stabilizes. Most instances in our experiments converge to the desired precision within $1000$ epochs.
    \item Early Stopping: To improve efficiency, we employ an early stopping criterion: if the change in energy between consecutive epochs falls below a predefined threshold ($10^{-4}$), we stop further training. This prevents unnecessary computations once the optimization has converged.
    \item Order Parameter Monitoring: At regular intervals, we compute the order parameter to track the physical characteristics of the state during training.
\end{enumerate}

For each experiment, we fix the circuit depth ($k$) and run multiple optimization instances with different initial parameters. Since the choice of initialization significantly influences convergence, we conduct approximately $100$ independent runs for statistical robustness. We repeat this process across different ansatze and the following models.
 
\begin{figure*}[t!]
    \centering
    \includegraphics[width=\linewidth]{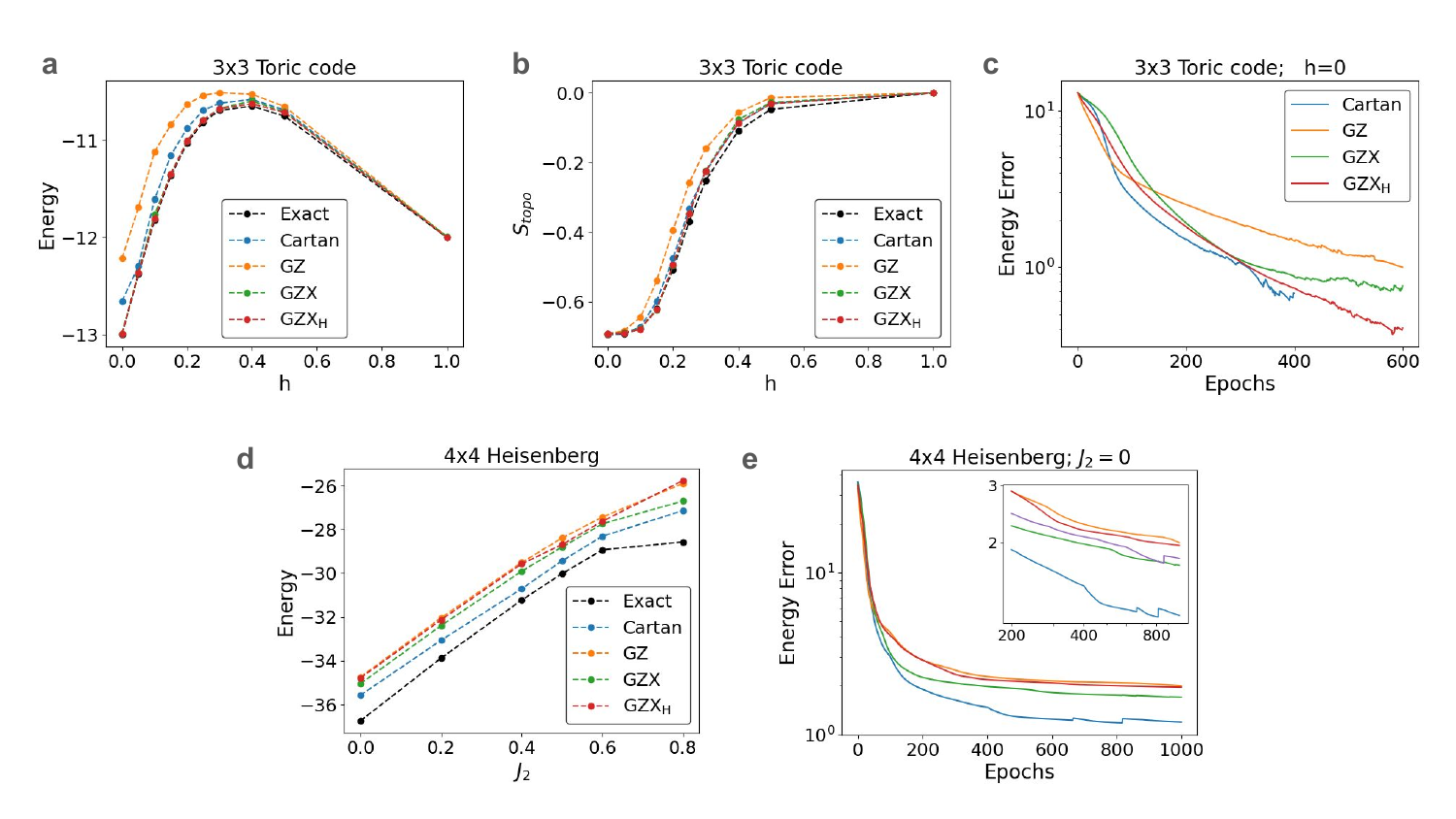}
    \caption{The figure presents results from our numerical simulations. We ran 100 instances for each Hamiltonian parameter ($J_2$ or $h$) and reported the average of the best-performing half among the converged results, plotted against the Hamiltonian parameter. The black dashed line shows the ground state energies obtained from exact diagonalization.
    For training plots, which depict the decay in energy error (difference between the obtained energy and the exact ground state energy) over epochs, we took the median energy across all instances at each epoch. Due to the early stopping criterion, different instances terminate at different points, leading to increased jaggedness towards the later epochs as fewer instances remain. The circuit depths were set to $k=3$ for the Heisenberg model and $k=4$ for the toric Code Hamiltonian.  
    (a) compares the converged energy values for different ansatze at various $h$ for a $3 \times 3$ toric Code Hamiltonian. (b) displays the converged topological entropy for different values of $h$. The GZX and ${\rm GZX}_{\rm H}$ ansatze consistently reach energies very close to the ground state for most instances, even outperforming the Cartan ansatz. (c) presents the variation of energy error with training epochs for toric code Hamiltonian. (d) shows the final converged energy for $4 \times 4$ Heisenberg model at various $J_2$ values. (e) presents the variation of energy error for the Heisenberg model with training epochs. 
    }
    \label{fig:application}
\end{figure*}

\subsection{Toric Code Model}

\begin{figure}
    \centering
    \includegraphics[width=\linewidth]{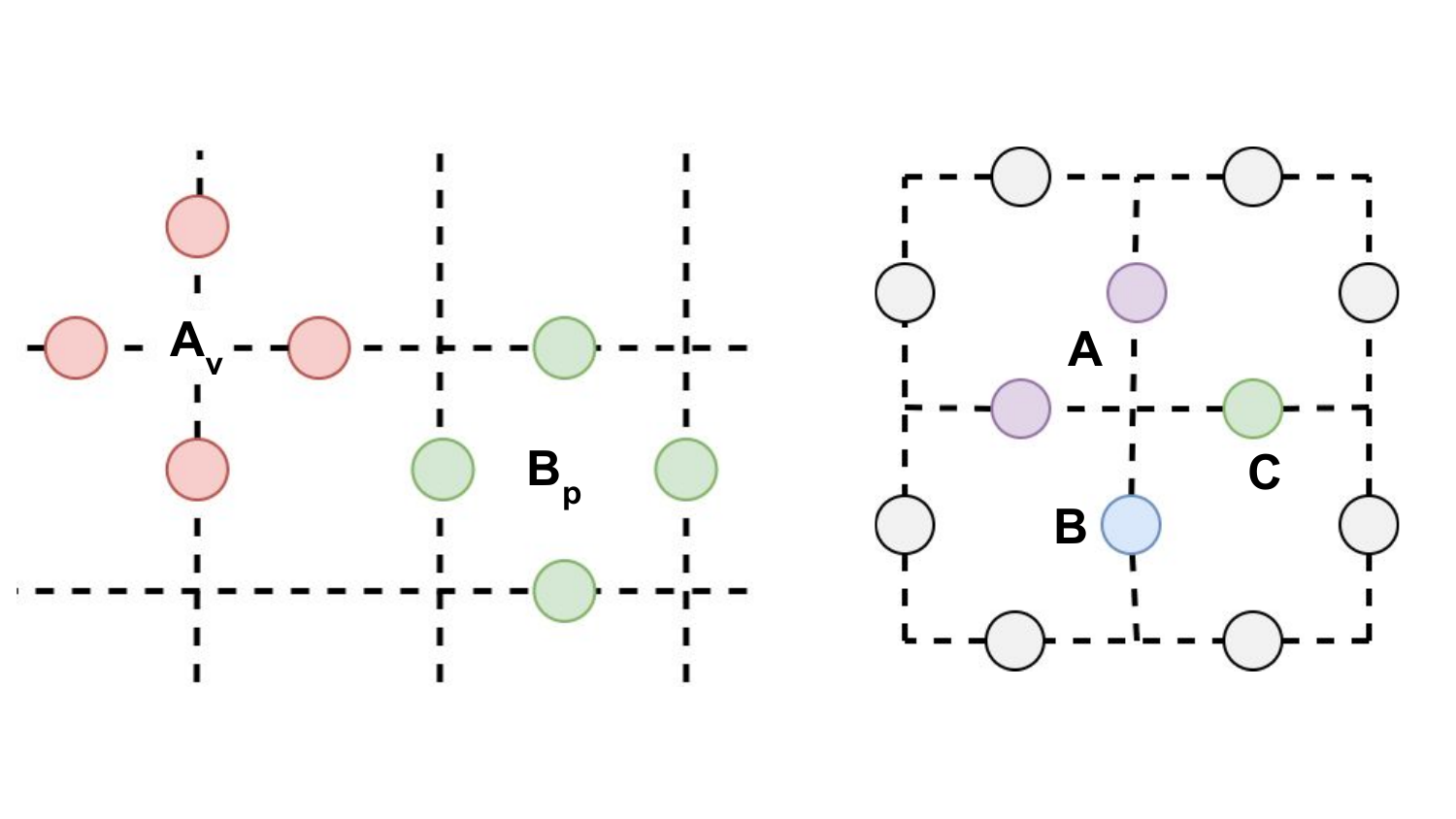}
    \caption{The left figure illustrates structures of the toric code model. The right figure indicates the subsystems involved in the computation of the topological entanglement entropy.}
    \label{fig:toric}
\end{figure}

The toric code Hamiltonian~\cite{KitaevToricCode} is a paradigmatic model in condensed matter physics and quantum information, playing an important role in error correction routines and is of interest to researchers as a substrate to exotic topological phases of matter. We study the generalized 2D toric code model,
\begin{equation}
    H = - (1 - h) \left[ \sum_v A_v + \sum_p B_p \right] - h \sum_{j=1}^N Z_j.
\end{equation}
Here, \( A_v \) and \( B_p \) are the vertex and plaquette operators, respectively, as illustrated in Figure~\ref{fig:toric}, left. These operators form the stabilizers of the model, with \( A_v \) acting on the qubits at the vertices of the lattice and enforcing local spin alignment, while \( B_p \) acts on the plaquettes and enforces spin-loop constraints. The inclusion of the Zeeman term, proportional to \( \sum_{j=1}^N Z_j \), introduces a tunable external magnetic field \( h \). This term breaks the perfect topological degeneracy of the ground state, rendering the model more physically realistic. Importantly, this modification creates a richer landscape of quantum phases and makes the Hamiltonian not easily simulated classically~\cite{satzinger2021realizing,chen2024sequential}. 

The Hamiltonian exhibits distinct phases as the parameter $h$ is varied~\cite{trebst2007breakdown,wu2012phase}. For small $h$, the system remains in a topologically ordered phase, characterized by long-range entanglement. In this regime, the $A_v$ and $B_p$ operators dominate, ensuring the ground state retains topological degeneracy. As $h$ increases, the Zeeman term becomes more significant, driving a transition to a trivial polarized phase where the spins align along the $Z$ direction. The critical point separating these phases marks a quantum phase transition, accompanied by a breakdown of topological order.

To quantify the presence of long-range entanglement, we use the topological entanglement entropy as the ``order parameter''. It is defined as the constant correction to the area law of subsystem entropies, i.e., for a subsystem $\mathcal{A}$ with area $L_{\mathcal{A}}$, it entropy scales as
\begin{equation}
    S(\mathcal{A}) = \alpha L_{\mathcal{A}} - \gamma,
\end{equation}
where a non-zero value of $\gamma$ indicates the presence of topological order~\cite{Levin2006,Kitaev2006}. In practice, $\gamma$ is extracted as a combination of the entropies for $3$ subsystems~\cite{Kitaev2006}, i.e.,
\begin{equation}
\begin{aligned}
        \gamma = S(\mathcal{A}) + S(\mathcal{B})& +  S(\mathcal{C}) +  S(\mathcal{ABC})\\
         -& S(\mathcal{AB})- S(\mathcal{AC})- S(\mathcal{BC}).
\end{aligned}
\end{equation}
Here the subsystem $\mathcal{A}$, $\mathcal{B}$, and $\mathcal{C}$ share boundaries in a particular way such that the entropy contributions from correlations across the boundaries explicitly cancel out, leaving only the topological entanglement entropy. The choice of the subsystems for the toric code model in our simulations is depicted in Figure~\ref{fig:toric}, right.

Figure~\ref{fig:application} (a,b) shows the energies and the topological entanglement entropies of the states found by our global gate ansatz, compared to the exact ground state of the toric code Hamiltonian. Most ansatze, except for GZ, converge very close to the exact ground state energy across the full range of Hamiltonian parameters. The ansatzes not only approximate the ground state energy but also correctly reproduce the topological entanglement entropy, indicating their ability to capture long-range entanglement. Among all ansatze tested, GZX and ${\rm GZX}_{\rm H}$ exhibit superior performance, even surpassing the Cartan ansatz. The rapid initial decay of the energy error suggests that these ansatze are highly trainable, meaning they provide large gradients and do not suffer from barren plateaus. The ability to sustain significant gradients throughout training ensures efficient learning.

\subsection{Heisenberg Model}

Next, to show the applicability of the proposed approach extends beyond what might seem like the highly curated ansatz that works only for the Toric code model, we explore the Heisenberg Hamiltonian, a widely studied model representing rich phase structures in magnetic systems. The specific Hamiltonian that we work with contains both the nearest and next-neighbor interaction terms~\cite{kochkov2021learning,nomura2021dirac,gong2014plaquette,choo2019two} on a square lattice, i.e.,
\begin{equation}
    H = \sum_{\langle i,j \rangle} \hat{S}_i \hat{S}_j + J_2 \sum{\langle \langle i,j \rangle \rangle} \hat{S}_i \hat{S}_j,
\end{equation}
where $\langle i,j \rangle$ and $\langle\langle i,j \rangle\rangle$ corresponds to indices of nearest and next-nearest neighbors respectively. The parameter $J_2$ controls the relative strength of next-nearest-neighbor interactions. Our simulation results [Figure~\ref{fig:application} (d,e)] demonstrates that the proposed global gate ansatze performs well for the Heisenberg model, achieving energy convergence close to the ground state across a wide range of parameters, showing its robustness beyond the Toric code model. These findings highlight the versatility of the approach in simulating complex quantum many-body systems. The energy errors in Figure~\ref{fig:application} (d) (deviation of the obtained energy from the exact energy) can be further reduced by increasing the training time: As indicated by the inset of Figure~\ref{fig:application} (e), the energy errors decay only as a power-law of the number of epochs, rather than a dramatic exponential slowing down.

\section{Discussion}

In this work, we developed a variational approach to harness the power of global quantum gates, native to many quantum computing platforms. Our proposed circuit ansatze consists of a finite number of global quantum gates interspersed with single-qubit rotations, enabling efficient implementation. We demonstrated the efficacy of this approach by using the global gate ansatze to prepare the ground states of the Heisenberg model and the Toric code model, achieving fast convergence and high accuracy. 

Our approach is not limited to the global control-Z gate as considered, but can be naturally generalized to other sets of global operations. Another promising direction is to investigate whether one can use a finite (or logarithm) number of global gates to simulate circuits that, with respect to local quantum gates, have logarithmic local depth but total linear depth. Such circuits are also barren plateau free but can potentially offer direct quantum advantage.

Although the scenario we studied, i.e., finite-depth (shallow) circuits with local measurements, can be efficiently classically simulated and hence do not offer \emph{direct} quantum advantages for statistics of local measurements alone, the post-training circuits can be ported as building blocks for subsequential tasks, for instance, preparing the physical target quantum states whose non-local correlations cannot be efficiently simulated. The circuit ansatze can be extended to other applications, such as unitary compiling or Hamiltonian discrimination using quantum machine learning~\cite{Leone2022}. It is crucial to emphasize that the high expressibility of the proposed global-gate ansatzes are irrespective of their classical simulatability with respect to local observables. Since shallow circuits, in general, do admit quantum advantages~\cite{Bravyi2018Quantum,Terhal2002Adaptive,Gao2017Quantum,Bermejo2018Architectures,Haferkamp2020Closing}, it is interesting to investigate whether the proposed ansatze can be adapted to these scenarios as well.

\vspace{8pt}
Acknowledgment.---The authors thank Marco Cerezo, Martin Larocca, Manas Sajjan, and Sabre Kais for helpful discussions. This work was supported in part by the U.S. Department of Energy, Office of Science, Office of Advanced Scientific Computing Research, through the Quantum Internet to Accelerate Scientific Discovery Program, and in part by the LDRD program at Los Alamos. V.S. also acknowledges support from the Quantum Computing Summer School at Los Alamos National Laboratory, and support from the U.S. Department of Energy Quantum Science Center.

\bibliography{ref}
\clearpage

\section*{Appendix}

\subsection{Frame Potential}
In the main text, we have used the trace distance between moments of the Haar random ensemble and the state ensemble generated by the circuit ansatze to quantify its expressibility. In this section we consider the expressibility measure induced by the Hilbert-Schmidt distance, 
\begin{equation}
A_{\rm HS}^{(t)}(\mathcal{C}) =  \left\| \int dU~ (U \rho U^{\dagger})^{\otimes t} - \int d\vec{\theta}~\left[U(\vec{\theta}) \rho U(\vec{\theta})^{\dagger}\right]^{\otimes t} \right\|_{\rm HS}.
\end{equation}
Interpretation of the parameters is the same Eq.~\ref{eq:distance} in the main text.

\begin{figure}[h!]\label{fig:framepotential}
    \includegraphics[width=\linewidth]{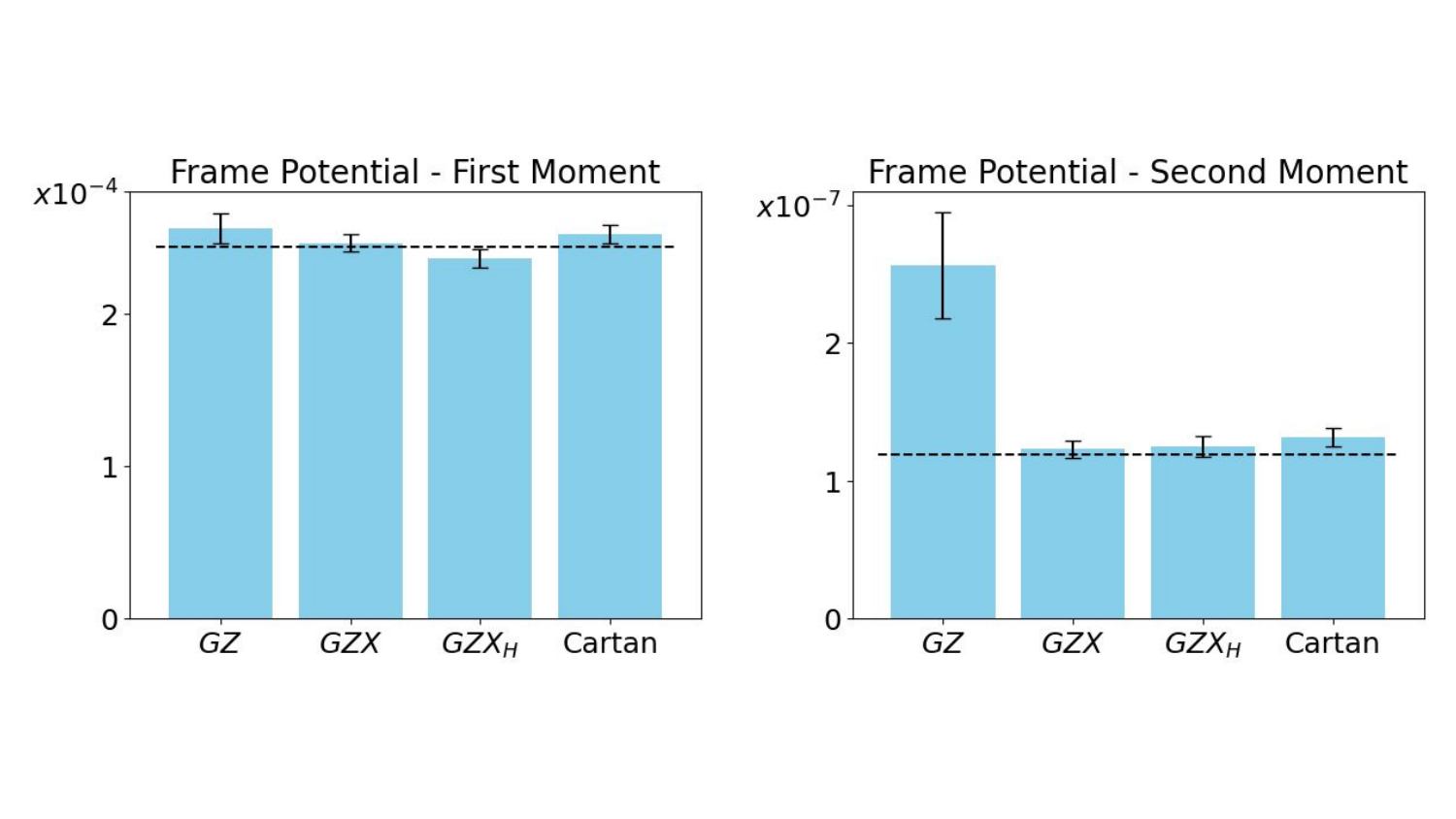}
    \caption{Here, we are comparing the expressibility of various 2D ansatze used for the toric code model. The figures show the first and second moment of frame potentials, respectively. The error bars are Standard error in mean estimation SEM = std / $\sqrt{S}$, where $S$ is the total number of samples used to estimate the moments. In our experiments, $S=4000$. The black dashed lines mark the frame potential for the Haar random distribution.}
\end{figure}

It can be shown that $A_{\rm HS}^{(t)}(\mathcal{C})$ is related to the frame potential, which, for ansatze $\mathcal{C}$, is defined as 
\begin{equation}
F^{(t)}_{\mathcal{C}} = \int\int |\langle\psi_\phi|\psi_\theta\rangle|^{2t} d\phi d\theta,
\end{equation}
where $|\psi_{\vec{\theta}} \rangle = U(\vec{\theta})|0\rangle$ is the output state generated by the ansatze at fixed parameters $\vec{\theta}$. The frame potential for the Haar random state ensemble, $F^{(t)}_{\rm Haar}$, can be defined similarly by replacing the above integral with that respect to the Haar measure. The relationship between the frame potentials and $A_{\rm HS}^{(t)}(\mathcal{C})$ is then:
\begin{equation}
F_{\mathcal{C}}^{(t)} - F_{\rm Haar}^{(t)} = A^{(t)}(\mathcal{C}) \geq 0,
\end{equation}
where the equality holds if and only if the ensemble of $|\psi_\theta\rangle$ forms a state $t$-design. Therefore, a smaller frame potential difference corresponds to higher expressibility. Figure~\ref{fig:framepotential} shows the frame potentials of the 2D ansatze for the toric code model as studied in the main text, confirming again their high expressibilities.

\subsection{Other ansatzes}

\begin{figure*}[t!]
    \centering
    \includegraphics[width=1\linewidth]{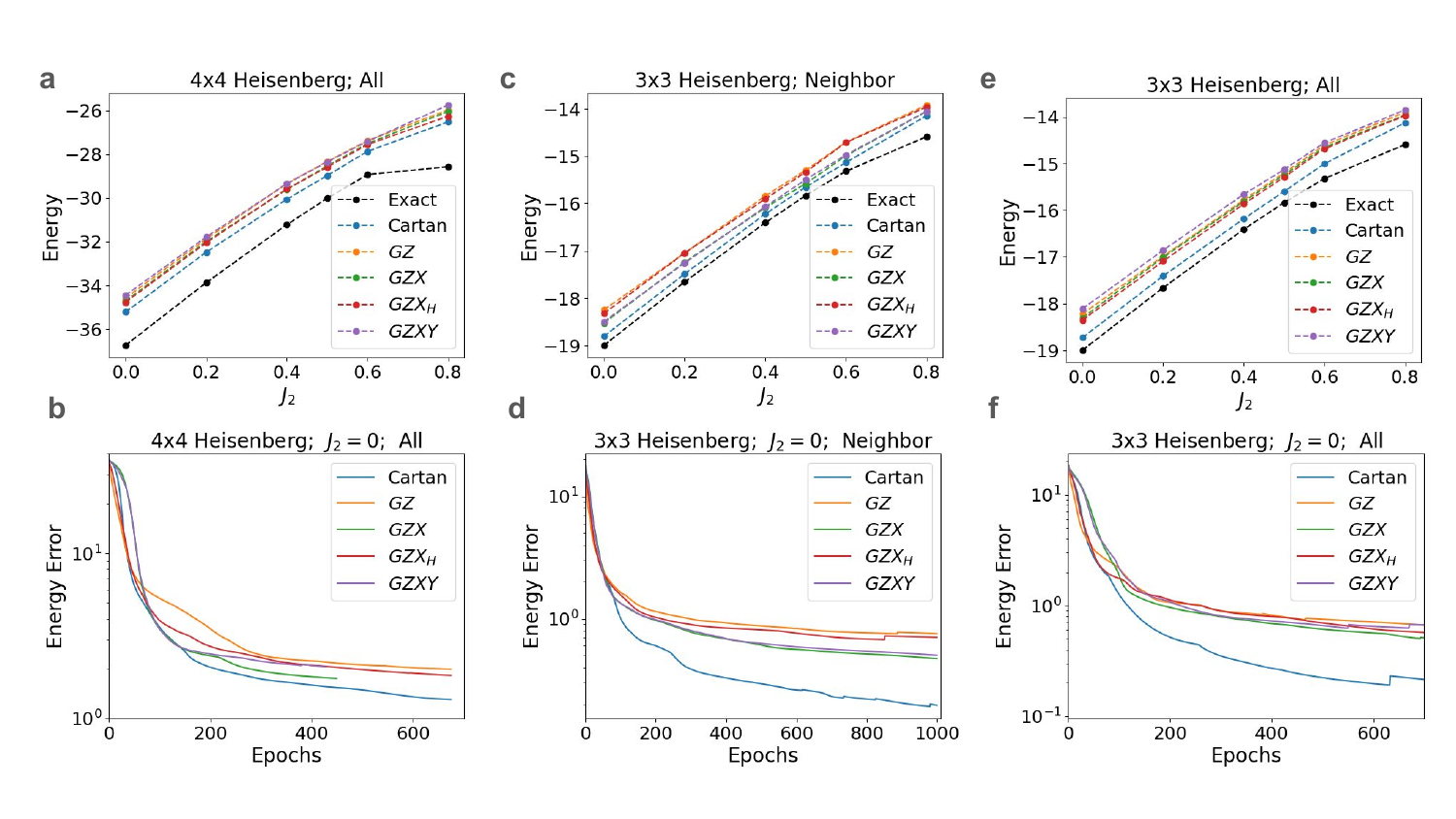}
    \caption{The figure presents results from our numerical simulations for the Heisenberg model at various system sizes for various types of ansatze. (a, c, e) show the converged ground state energies. (b, d, f) present the variation of energy error with training epochs. Training parameters are the same as described in Figure~\ref{fig:application} in the main text.
    \label{fig:ap:Heisenberg}
    }
\end{figure*}

The 2D ansatze used on the square lattice Heisenberg model shown in the main text is called a `Neighbor'-shaped ansatze, as for a given plaquette, only the neighboring edges are connected. 
We also experimented with another ansatz structure, which has all-to-all connectivity among the qubits in a plaquette, and we called it the `All' ansatz. Due to the dense connectivity in the ansatz, we expect it to have more expressibility and higher representation power for correlations. On the other hand, the values of gradients would be smaller, and this circuit is expected to be slightly harder to train. 
Figure~\ref{fig:ap:Heisenberg} shows the results obtained for training using the `All'-ansatz, as well as the `Neighbor'-shaped ansatze for various system sizes.

\subsection{Trainabiilty}
\begin{figure*}[b!]
    \centering
    \includegraphics[width=1\linewidth]{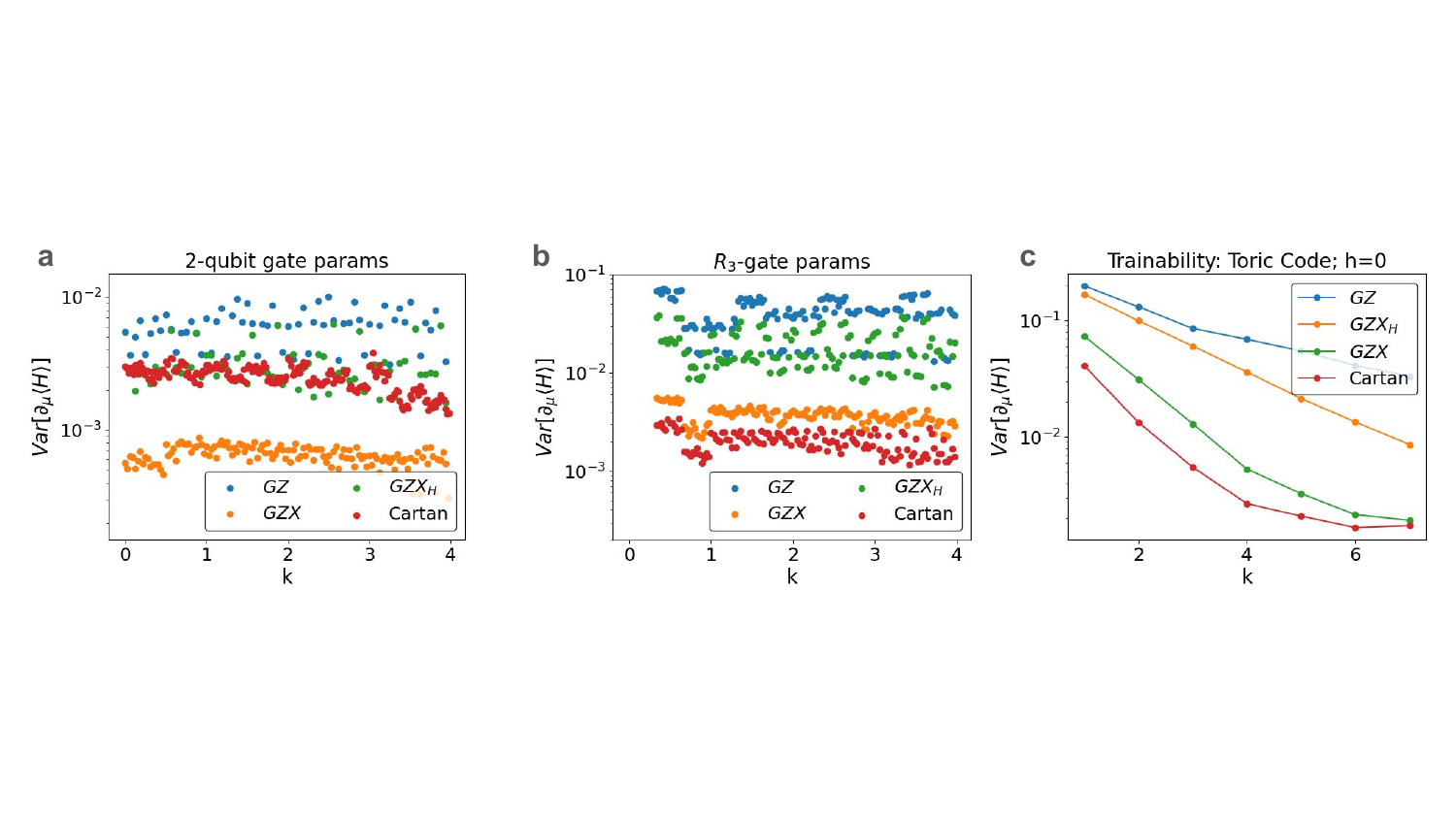}
    
    \caption{These plots show the variance of gradients with respect to all parameters in the ansatze for the toric code Hamiltonian, with circuit depth $k = 4$. The $y$-axis represents $\text{Var}[\partial_\mu \langle H \rangle]$ for various parameters $\mu$, and the $x$-axis lists all circuit parameters in order of application. Circuit depth \( k \) is marked by integers, with parameters of each layer positioned between consecutive \( k \) values (e.g., points between \( k = 1 \) and \( k = 2 \) correspond to the second layer). (a) Displays the gradient variance for parameters in the \( R_3 \)-gate layers. Since $R_3 = R_Z(\theta_3) R_Y(\theta_2) R_Z(\theta_1)$, each layer has \( 3N \) parameters, leading to \( 3N \) data points per layer (where \( N \) is the system size in number of qubits). The gradients with respect to the first \( R_z \) gates are zero and are, therefore, not visible in the plot.  (b) Shows \( \text{Var}[\partial_\mu \langle H \rangle] \) for parameters in the two-qubit gates, namely $CZ$, $R_{XX}$, $R_{YY}$, and $R_{ZZ}$, each having a single parameter. The number of data points varies across different ansatze because they contain different numbers of two-qubit parameters. Specifically, GZ and \( \rm{GZX}_H \) have \( M \) parameters per layer, GZX has \( 2M \) parameters per layer, and Cartan has \( 3M \) parameters per layer, where \( M \) is the number of two-qubit gate locations in the lattice. For the \( 3 \times 3 \) toric code (\( N = 12 \)), \( M = 16 \) (see Figure~\ref{fig:ansatze} in the main text). (c) Shows the change in \( \text{Var}[\partial_\mu \langle H \rangle] \) with increasing circuit-depth for a \( 3 \times 3 \) toric code Hamiltonian at h=0.} 
    \label{Fig:Trainability}.  
\end{figure*}

\paragraph{Choice of $\mu$}:
We computed the gradient variance with respect to all parameters to determine the choice of $\mu$ as a marker to compare gradients across various ansatze. Figure~\ref{Fig:Trainability} illustrates the ${\rm Var[}\partial_\mu \langle H \rangle]$ values across all parameters in different ansatze. As shown in the figure, the gradients for a given ansatz tend to cluster within a specific range, indicating that any parameter choice would be equally valid for trainability comparisons. Our specific choice of $\mu$, the last $R_Y$ gate of the first $R_3$-gate layer in each ansatz, was inspired by \cite{zhang2024absence}. 

\paragraph{Toric code}:
In the main text, we present trainability data only for a simplified \( Z_N \) Hamiltonian. However, we also computed gradient variances for other Hamiltonians using various ansatze. Figure~\ref{Fig:Trainability} illustrates how the gradient variance changes with increasing gate depth for the toric code Hamiltonian with 12 qubits.

\end{document}